\documentclass{article}

\usepackage{latexsym}

\newcommand{\sbs}{\subseteq} \newcommand{\Eqv}{\Longleftrightarrow}
\newcommand{\eqv}{\Leftrightarrow} 
\newcommand{\dam}{\Diamond} \newcommand{\sat}{{\models}}
\newcommand{\dar}{{\downarrow}}

\newcommand{\R}{{\bf R}}

\newcommand{\ptimes}{\dot{\times}}

\newcommand{\sk}{{\sf K}}\newcommand{\sa}{{\sf A}}
\newcommand{\sfl}{{\sf L}}

 \newcommand{\cl}{{\mathcal L}}
\newcommand{\cb}{{\mathcal B}} \newcommand{\ct}{{\mathcal T}}
\newcommand{\cm}{{\mathcal M}} \newcommand{\cp}{{\mathcal P}}
\newcommand{\cf}{{\mathcal F}} \newcommand{\cg}{{\mathcal G}}
 \newcommand{\co}{{\mathcal O}}
 \newcommand{\clc}{{\mathcal C}}
\newcommand{\clr}{{\mathcal R}}

\newcommand{\Rem}[2]{{\sf Rem}^{#1}#2}
\newcommand{\tsat}{\sat_{\ct}}
\newcommand{\bsat}{\sat_{\cb}}
\newcommand{\msat}{\sat_{\cm}}
\newcommand{\csat}{\sat_{\clc}}
\newcommand{\notcsat}{\not\sat_{\clc}}
\newcommand{\bland}{\bigwedge} \newcommand{\blor}{\bigvee}
\newcommand{\todam}{R_{\Box}}
\newcommand{\tol}{R_{\sf\scriptstyle K}}

\newcommand{\MP}{\bf MP^*}
\newcommand{\MPR}{\vdash_{\bf\scriptstyle MP^*}}
\newcommand{\A}{{\forall}}
\newcommand{\E}{{\exists}}

\newtheorem{theorem}{\bf Theorem}
\newtheorem{lemma}[theorem]{\bf Lemma}
\newtheorem{prop}[theorem]{\bf Proposition}
\newtheorem{corol}[theorem]{\bf Corollary}

\newenvironment{dfn}{\par\medskip\addtocounter{theorem}{1}%
  \noindent{\bf Definition \arabic{theorem}}\quad}{\medskip}

\newcommand{\qed}{\vrule height5pt width3pt depth0pt}

\newenvironment{prf}{\noindent {\sc Proof.}}{
{\nobreak\hfill \qed \par \medbreak}}

\title{\sf Modal Logics of Topological Spaces\\
{\rm \small (Extended abstract of doctoral dissertation)}}

\author{ Konstantinos Georgatos\thanks{Department
of Mathematics, Graduate School and University Center,  City University of
New York,  33 West 42nd Street, New York, NY 10036 }}

\date{May 3, 1993}

\begin{document}

\maketitle

\section{Introduction}

In this thesis we shall present two logical systems, $\bf MP$ and
$\MP$, for the purpose of
reasoning about {\em knowledge} and {\em effort\/}. These logical systems  will
be interpreted in a spatial context and therefore, the abstract concepts of
knowledge and effort will be defined by concrete mathematical concepts.

Our general framework consists of a set of possible worlds (situations,
scenarios, consistent theories, etc.) A {\em state of knowledge} is a subset of
this set and our knowledge consists of all facts common to the worlds belonging
to this subset. This subset of possibilities can be thought as our {\em view}.
Thus two knowers having distinct views can have different knowledge. This
treatment of knowledge agrees with the traditional one
(\cite{HI}, \cite{HAM}, \cite{PR}, \cite{CM}, \cite{FHV})
expressed in a variety of contexts (artificial intelligence, distributed
processes, economics, etc.)

Our treatment is based on the following simple observation
\begin{quote}
``a restriction of our view increases our knowledge.''
\end{quote}
This is because a smaller set of possibilities implies a greater amount of
common facts. Moreover, such a restriction can only be possible due to an
increase of information. And such an information increase can happen with
spending of time or computation resources. Here is where the notion of effort
enters. A restriction of our view is {\em dynamic} (contrary to the view itself
which is a state) and is  accompanied by effort during which a greater amount
of information becomes available to us (Pratt expresses a similar idea in the
context of processes \cite{PR92}.)

We make two important assumptions.

Our knowledge has a subject. We collect information for a specific purpose.
Hence we are not considering arbitrary restrictions to our view but restrictions
parameterized by possibilities contained in our view, i.e. neighborhoods of
possibilities. After all, only one of these possibilities is our actual
state. This crucial assumption enables us to express topological
concepts and use a mathematical set-theoretic setting as semantics. Without
such an assumption these ideas would have been expressed in the familiar theory
of intuitionism (~\cite{HEY56},~\cite{DU77},~\cite{TRO88}.) As Fitting points
out in~\cite{FIT69}
\begin{quote}
``Let $\langle\cg,\clr,\sat\rangle$ be a [intuitionistic, propositional] model.
$\cg$ is intended to be a collection of possible universes, or more properly,
states of knowledge. Thus a particular $\Gamma$ in $\cg$ may be considered as a
collection of (physical) facts known at a particular time. The relation $\clr$
represents (possible) time succession. That is, given two states of knowledge
$\Gamma$ and $\Delta$ of $\cg$, to say $\Gamma\clr\Delta$ is to say: if we now
know $\Gamma$, it is possible that later we will know $\Delta$.''
\end{quote}
Considering neighborhoods and, inevitably, points which parameterize
neighborhoods, the important duality between the facts, which constitute our
knowledge, and the possible worlds, where such facts hold, emerges.

The other assumption is that of indeterminacy. Each state of knowledge is closed
under logical deduction. Thus an increase of knowledge can happen only by a
piece of evidence or information given from outside. Our knowledge is {\em
external} (a term used by Parikh to describe a similar idea in~\cite{PAR87}.)
This fact leads to indeterminacy (we do not know which kinds of information will
be available to us, if at all) and resembles indeterminacy expressed in
intuitionism through the notion of {\em lawless} sequence
(see~\cite{KR58},~\cite{TRO77}) where, not surprisingly, topological notions
arise.

To illustrate better these simple but fundamental ideas we present the following
examples:
\begin{itemize}
\item Suppose that a machine emits a stream of binary digits representing the
output of a recursive function $f$. After time $t_1$  the machine emitted
the stream $111$. The only information we have about the function being computed
at this time on the basis of this (finite) observation is that
$$f(1)=f(2)=f(3)=1.$$
As far as our knowledge concerns $f$ is indistinguishable from the constant
function ${\bf 1}$, where ${\bf 1}(n)=1$ for all $n$. After some additional time
$t_2$, i.e. spending more time and resources, $0$ might appear and thus we could
be able to distinguish $f$ from $\bf 1$.
In any case, each binary stream will be an initial segment of $f$ and this
initial segment is a neighborhood of $f$. In this way, we can acquire more
knowledge for the function the machine computes. The space of finite binary
streams is a structure which models computation. Moreover, this space comprises
a topological space. The set of binary streams under the prefix ordering is
an example of {\em Alexandrov topology} (see~\cite{V}.)

\item A policeman measures the speed of passing cars  by means of a device. The
speed limit is $80$ km/h. The error in measurement which the device introduces
is $1$ km/h. So if a car has a speed of $79.5$ km/h and his device measures
$79.2$ km/h then he knows that the speed of the passing car lies in the interval
$(78.2,80.2)$ but he does not know if the car exceeds the speed limit because
{\em not all} values in this interval are more than $80$. However, measuring
again and combining the two measurements or acquiring a more accurate device he
has the possibility of knowing that a car with a speed of $79.5$ km/h does not
exceed the speed limit. Note here that if the measurement is, indeed, an open
interval of real line and the speed of a passing car is exactly $80$ km/h then
he would never know if such a car exceeded the speed limit or not.

\end{itemize}

To express this framework we use two modalities $\sk$ for knowledge and $\Box$
for effort. Moss and Parikh observed in~\cite{MP} that if the formula
$$
A\to\dam\sk A
$$
is valid, where $A$ is an atomic predicate and $\dam$ is the dual of the $\Box$,
i.e. $\dam\equiv\neg\Box\neg$, then the set which $A$ represents is an open set
of the topology where we interpret our systems.
Under the reading of $\dam$ as ``possible'' and $\sk$ as ``is known'', the above
formula says that
\begin{quote}
``if $A$ is true then it is for $A$ possible to be known'',
\end{quote}
i.e. $A$ is {\em affirmative}. Vickers defines similarly an
affirmative assertion  in~\cite{V}
\begin{quote}
``an assertion is affirmative iff it is true precisely in the circumstances when
it can be affirmed.''
\end{quote}
The validity of the dual formula
$$
\Box\sfl A\to A,
$$
where $\sfl$ is the dual of $\sk$, i.e. $\sfl\equiv\neg\sk\neg$, expresses the
fact that the set which $A$ represents is closed, and hence $A$ is {\em
refutative}, meaning if it does not hold then it is possible to know that. The
fact that
affirmative and refutative assertions are represented by opens and closed
subsets, respectively, should not come to us as
a surprise. Affirmative assertions are closed under infinite disjunctions and
refutative assertions are closed under infinite conjunctions. Smyth
in~\cite{SMY83} observed first these properties in semi-decidable properties.
Semi-decidable properties are those properties whose truth set is r.e. and are a
particular kind of affirmative assertions. In fact, changing our power of
affirming or computing we get another class of properties with a similar
knowledge-theoretic character. For example, using polynomial algorithms
affirmative assertions become polynomially semi-decidable properties. If an
object has this property then it is possible to know it with a polynomial
algorithm even though it is not true we know it now.

Does this framework suffers from the problem of logical omniscience? Only in
part. Expressing effort we are able to bound the increase of knowledge depending
on information
(external knowledge.) Since the modality $\sk$ which corresponds to knowledge is
axiomatized by the normal modal logic of $\bf S5$, knowledge is closed under
logical deduction. However, because of the strong computational character of
this framework it does not seem unjustified to assume that in most cases
(as in the binary streams example) a finite amount of data restricts our
knowledge to a finite number of (relevant) formulae. Even without such an
assumption we can incorporate the effort to deduce the knowledge of a property
in the passage from one state of knowledge to the other.

We have made an effort to present our material somewhat independently. However,
knowledge of basic modal logic, as in \cite{CH}, \cite{HC}, or~\cite{FIT93},
is strongly recommended.

The language and semantics of our logical framework is presented in
Chapter~\ref{ch:systems}. In the same Chapter we present two systems: $\bf MP$
and $\MP$. The former was introduced in~\cite{MP} and was proven complete for
arbitrary sets of subsets. It soon became evident that such sets of subsets
should be combined, whenever it is possible, to yield a further increase of
knowledge or we should assume a previous state of other states of knowledge
where such states are a possible. Therefore the set of subsets should
be closed under union and intersection. Moreover, topological notions
expressed in $\bf MP$ make sense only in topological models. For this reason we
introduce
an extension of the set of axioms of $\bf MP$ and we call it $\MP$. In
Chapter~\ref{ch:sem}, we study the topological models of $\MP$ by semantical
means. We are able to prove the reduction of the theory of topological models to
models whose associated set of subsets is closed under finite union and
intersection. Finding for each satisfiable formula a model of bounded size we
prove decidability for $\MP$. The results of this chapter will appear
in~\cite{KG1}. In Chapter~\ref{ch:complete}, we prove that $\MP$
is a complete system for topological models as well as topological models
comprised by closed subsets. We also give necessary and sufficient conditions
for turning a Kripke frame into such a topological model. In
Chapter~\ref{ch:algebraic}, we present the modal algebras of $\bf MP$ and $\MP$
and some of their properties. Finally, in Chapter~\ref{ch:future}, we present
some of our ideas towards future work.

\section{Two Systems: $\bf MP$ and $\MP$}
\label{ch:systems}

In section~\ref{sec:lang} we shall present a language and
semantics which  appeared first in \cite{MP}. In section~\ref{sec:mp},
we shall present the axiom system $\bf MP$, introduced and proven
sound and complete with a class of models called {\em subset spaces} in
\cite{MP}, and the axiom system $\MP$, introduced by us, which we
shall
prove sound and complete for, among other classes, the class of topological
spaces.

\subsection{Language and Semantics}
\label{sec:lang}

We follow the notation of \cite{MP}.

Our language is bimodal and propositional.
Formally, we start with a countable
set $\sa$ of
{\em atomic formulae} containing two distinguished elements $\top$ and $\bot$.
Then the {\em language} $\cl$ is the
least set such that $\sa\sbs\cl$ and closed under the following rules:

$$\frac{\phi,\psi\in\cl}{\phi\land\psi\in\cl}\qquad
\frac{\phi\in\cl}{\neg\phi,\Box\phi,\sk\phi\in\cl}$$

\medskip
The above language can be interpreted inside any spatial context.

\begin{dfn}
Let $X$ be a set and $\co$ a subset of the powerset of $X$, i.e.
$\co\sbs\cp(X)$ such that $X\in\co$.
We call the pair $\langle X,\co \/\rangle$ a {\em subset space}.
A {\em model} is a triple $\langle X,\co,i\/\rangle$, where
$\langle X,\co \rangle$ is a subset space and $i$ a map from $\sa$
to $\cp(X)$ with $i(\top)=X$ and $i(\bot)=\emptyset$ called {\em
initial interpretation}.
\end{dfn}

We denote the set
$\{ (x,U) : x\in X , U\in\co , \hbox{ and } x\in U\}\sbs X\times\co$
by $X\dot{\times}\co$. For each $U\in\co$ let $\dar U$ be the set
$\{ V : V\in\co \hbox{ and } V\sbs U \}$  the lower closed set
generated by $U$ in the partial order $(\co,\sbs)$,
i.e. $\dar U=\cp(U)\cap \co$.

\begin{dfn}
The {\em satisfaction relation} $\msat$, where $\cm$ is the model
$\langle X,\co,i\/\rangle$,
is a subset of $(X\dot{\times}\co)\times\cl$ defined recursively  by
(we write $x,U\msat\phi$ instead of $((x,U),\phi)\in\msat$):
$$
\begin{array}{ll}
x,U\msat A & \hbox{iff}\quad x\in i(A),\hbox{ where }A\in\sa \medskip
\\ x,U\msat \phi\land\psi & \hbox{if}\quad x,U\msat\phi \hbox{ and }
                                           x,U\msat \psi\medskip  \\
x,U\msat \neg\phi & \hbox{if}\quad x,U \not\msat\phi        \medskip
\\ x,U\msat \sk\phi & \hbox{if}\quad\hbox{for all } y\in U,\quad
                                           y,U\msat\phi \medskip  \\
x,U\msat \Box\phi & \hbox{if} \quad\hbox{for all } V\in\dar U \hbox{
such that } x\in V, \quad x,V\msat\phi.
\end{array}
$$
If $x,U\msat\phi$ for all $(x,U)$ belonging to $ X\dot{\times}\co$
then $\phi$ is {\em valid} in $\cm$, denoted by $\cm\sat\phi$.
\end{dfn}

We abbreviate $\neg\Box\neg\phi$ and $\neg\sk\neg\phi$ by
$\dam\phi$ and $\sfl\phi$ respectively.
We have that
$$
\begin{array}{ll}
x,U\msat \sfl\phi & \hbox{if there exists } y\in U \hbox{ such that }
                                             y,U\msat\phi \medskip \\
x,U\msat \dam\phi & \hbox{if there exists } V\in\co \hbox{ such that }
V\sbs U,\  x\in V,\hbox { and } x,V\msat\phi.
\end{array}
$$

Many topological properties are expressible in this logical
system in a natural way. For instance, in a model where the subset
space is a topological space,  $i(A)$ is  {\em open} whenever
$A\to\dam\sk A$ is valid in this model. Similarly, $i(A)$ is {\em nowhere
dense} whenever $\sfl\dam\sk\neg A$ is valid (cf. \cite{MP}.)

\medskip
\noindent{\em Example\/.}
Consider the set of {\em real numbers} $\R$ with the usual topology of
open intervals. We define the following three
predicates:\begin{eqnarray*}
{\tt pi} & \hbox{where} & i({\tt pi}) = \{ \pi \} \\
{\tt I}_1 & \hbox{where} & i({\tt I}_1)=(-\infty,\pi ]\\
{\tt I}_2 & \hbox{where}  & i({\tt I}_2)=(\pi,+\infty)\\
{\tt Q} & \hbox{where} & i({\tt Q})=\{ q : q\hbox{ is rational }\}.
\end{eqnarray*}

There is no real number $p$ and open set $U$ such that
$p,U\sat\sk {\tt pi}$ because that would imply $p=\pi$ and
$U=\{\pi\}$ and there are no singletons which are open.

A point $x$ belongs to the {\em closure} of a set $W$ if every
open $U$ that contains $x$ intersects $W$.
Thus $\pi$ belongs to the closure of $(\pi,+\infty)$, i.e every open
that contains $\pi$ has a point in $(\pi,+\infty)$. This means that
for all $U$ such that $\pi\in U$, $\pi,U\sat\sfl {\tt I}_2$,
therefore $\pi,\R\sat\Box\sfl {\tt I}_2$. Following
the same reasoning $\pi,\R \sat\Box  {\tt I}_1$, since $\pi$
belongs to the closure of $(-\infty,\pi ]$.

A point $x$ belongs to the {\em boundary} of  a set $W$ whenever
$x$ belong to the closure of $W$ and $X-W$. By the above, $\pi$
belongs to the boundary of $(-\infty,\pi ]$ and
$\pi,\R \sat \Box ( \sfl {\tt I}_1 \land \sfl {\tt I}_2 )$.

A set $W$ is {\em closed} if it contains its closure. The interval
$i({\tt I}_1)=(-\infty,\pi ]$ is closed and this means that the
formula $\Box\sfl {\tt I}_1 \to {\tt I}_1$ is valid.

A set $W$ is {\em dense} if all opens contain a point of $W$. The set
of rational numbers is dense which translates to the fact that the
formula $\Box\sfl {\tt Q}$ is valid. To exhibit the reasoning in this
logic, suppose that the set of rational numbers was closed then both
$\Box\sfl {\tt Q}$ and $\Box\sfl {\tt Q} \to {\tt Q}$ would be valid.
This implies that ${\tt Q}$  would be valid which means that all reals
would be rationals. Hence the set of rational numbers is not closed.
\medskip

\subsection{$\bf MP$ and $\MP$}
\label{sec:mp}
The
axiom system $\bf MP$ consists of axiom schemes \ref{ax:prop}
through~\ref{ax:boxk} and rules of  table~\ref{table:mp}
(see page~\pageref{table:mp}) and appeared first in \cite{MP}.

\begin{table}
\begin{center}
\fbox{
\begin{minipage}{5in}
\vspace{.3in}
\noindent{\bf Axioms}
\begin{enumerate}
\item All propositional tautologies \label{ax:prop}
\item $(A\to\Box A) \land (\neg A\to\Box\neg A)$, for
$A\in\sa$\label{ax:atom}
\item
$\Box(\phi\to\psi)\to(\Box\phi\to\Box\psi)$\label{ax:normb}
\item $\Box\phi\to\phi$
\item $\Box\phi\to\Box\Box\phi$\label{ax:s4}
\item $\sk(\phi\to\psi)\to(\sk\phi\to\sk\psi)$\label{ax:normk}
\item $\sk\phi\to\phi$
\item $\sk\phi\to\sk\sk\phi$
\item $\phi\to\sk\sfl\phi$\label{ax:s5}
\item $\sk\Box\phi\to\Box\sk\phi$\label{ax:boxk}
\item $\dam\Box\phi\to\Box\dam\phi$\label{ax:direct}
\item $\dam(\sk\phi\land\psi)
       \land\sfl\dam(\sk\phi\land\chi)
       \to\dam(\sk\dam\phi\land\dam\psi\land\sfl\dam\chi)$\label{ax:union}
\end{enumerate}

\noindent{\bf Rules}

$$\frac{\phi\to\psi,\phi}{\psi}\ \hbox{\footnotesize MP}$$
$$\frac{\phi}{\sk\phi}\ \hbox{\footnotesize
\footnotesize \sk-Necessitation} \qquad
  \frac{\phi}{\Box\phi}\ \hbox{\footnotesize $\Box$-Necessitation}$$
\medskip
\end{minipage}
}
\end{center}
\bigskip
\caption{\label{table:mp} Axioms and Rules of $\MP$}
\end{table}

The following was proved in \cite{MP}.
\begin{theorem}
The axioms and rules of $\bf MP$ are sound and complete with respect to subset
spaces.
\end{theorem}

We add the axioms \ref{ax:direct} and~\ref{ax:union} to form the system $\MP$
for the purpose
of  axiomatizing spaces closed under union and intersection and, in
particular, topological spaces.

A word about the axioms (most of the following
facts can be
found in any introductory book about modal logic, e.g. \cite{CH} or \cite{GL}.)
The axiom~\ref{ax:atom}  expresses the fact that the truth of atomic formulae
is independent of the choice of subset and depends only on the choice of point.
This is the first example of a class of formulae which we are going to call
{\em bi-persistent} and their identification is one of the key steps to
completeness. Axioms \ref{ax:normb} through
\ref{ax:s4} and axioms \ref{ax:normk} through \ref{ax:s5} are used to axiomatize
the normal modal logics {\bf S4} and {\bf S5} respectively. The former group of
axioms expresses the fact that the passage from one subset to a restriction of
it is done in a constructive way as actually happens to an increase of
information or a spending of resources (the classical
interpretation of necessity in intuitionistic logic is axiomatized in the
same way). The latter group is generally used for axiomatizing logics of
knowledge.

Axiom~\ref{ax:boxk} expresses the fact that if a formula holds in arbitrary
subsets is going to hold as well in the ones which are neighborhoods of a
point. The converse is not sound.

Axiom~{\ref{ax:direct}} is a well-known formula which characterizes
{\em incestual} frames, i.e. if two points $\beta$ and $\gamma$ in a
frame can be accessed
by a common point $\alpha$ then there is a point $\delta$ which
can be accessed by both $\beta$ and $\gamma$. It
appeared in the equivalent form (in \cite{MP})
$$\dam\Box\phi\land\dam\Box\psi\to\dam\Box(\phi\land\psi)$$
and was proved sound in
subset spaces closed under (finite) intersection.

Obviously our attention is focused on axiom~\ref{ax:union}. It is sound in
spaces closed under (finite) union and intersection as the following proposition
shows.

\begin{prop}
Axioms 1 through \ref{ax:union} are sound in the class of subset spaces closed
under finite union and intersection. \end{prop}

\begin{prf}
Soundness for Axioms 1 through 11 is easy.
For Axiom~\ref{ax:union}, suppose
$$x,U\sat\dam(\sk\phi\land\psi)
       \land\sfl\dam(\sk\phi\land\chi).$$

Since $x,U\sat\dam(\sk\phi\land\psi)$, there exists
$U_x\sbs U$ such that
$$x,U_x\sat\sk\phi\land\psi$$
 and, since
$x,U\sat\sfl(\dam\sk\phi\land\Box\chi)$, there exists $y\in U$ and
$U_y\sbs U$ such that
$$y,U_y\sat\sk\phi\land\chi.$$
We now have that $U_x\cup
U_y\sbs U$ (we assume closure under unions.)
Thus
$$x,U_x\cup U_y\sat\sk\dam\phi,\quad y,U_x\cup U_y\sat\sk\dam\phi,
\quad x,U_x\cup U_y\sat\dam\psi,\quad \hbox{and}\quad y,U_x\cup
U_y\sat\dam\chi.$$
Therefore, $$x,U\sat\dam(\sk\dam\phi\land\dam\psi\land\sfl\dam\chi).$$
\end{prf}

With the help of axiom~\ref{ax:union}  we are able to prove
the key lemma~\ref{lem:main} which leads to the DNF Theorem (page
\pageref{thm:dnf}.)
and this is the only place where we actually use it. Any formula,  sound in the
class of subset spaces closed under finite union and intersection, which
implies the formula (note the difference from axiom~\ref{ax:union})
$$\dam(\sk\phi\land\psi)\land\sfl\dam(\sk\phi\land\chi)
\to\dam(\sk\phi\land\psi\land\sfl\chi)$$
where $\dam\phi\to\Box\phi$, $\dam\psi\to\psi$ and $\chi\to\Box\chi$ are
theorems, can replace axiom~\ref{ax:union}.


\section{A Semantical analysis of $\MP$}
\label{ch:sem}

\subsection{Stability and Splittings}
\label{sec:stability}
Suppose that $X$ is a set and $\ct$ a topology on $X$.  In the
following we assume that we are working in the topological space
$( X,\ct )$. Our aim  is to find a partition of $\ct$, where a given
formula $\phi$ ``retains its truth value'' for each point throughout a
member of this partition. We shall show that there exists a finite
partition of this kind.

\begin{dfn}
Given a finite family $\cf=\{ U_1,\ldots,U_n\}$ of opens, we define
the {\em remainder} of (the principal ideal in $(\ct,\sbs)$ generated
by) $U_k$ by
$$\Rem{\cf}{U_k}\quad=\quad\dar U_k - \bigcup_{U_k\not\subseteq U_i}\dar U_i.$$
\end{dfn}

\begin{prop} In a finite set of opens $\cf=\{U_1,\ldots,U_n\}$ closed
under intersection, we have
$$\Rem{\cf}{U_i}\quad=\quad\dar U_i - \bigcup_{U_j\subset U_i}\dar U_j,$$
for $i=1,\ldots,n$. \end{prop}

We denote $\bigcup_{U_i\in\cf}\dar U_i$ with $\dar\cf$.

\begin{prop} If $\cf=\{U_1,\ldots,U_n\}$ is a finite family of opens,
closed under intersection, then
\label{prop:part}
\renewcommand{\theenumi}{\alph{enumi}}
\begin{enumerate}
\item
$\Rem{\cf}{U_i}\cap\Rem{\cf}{U_j}=\emptyset$, for $i\not= j$,
\item
$\bigcup^n_{i=1}\Rem{\cf}{U_i}=\dar\cf$,
i.e.
$\{\Rem{\cf}{U_i}\}^n_{i=1}$ is a partition of $\dar\cf$. We call
such an $\cf$ a {\em finite splitting (of $\dar\cf$\/)\/},
\item
if $V_1,V_3\in\Rem{\cf}{U_i}$ and $V_2$ is an open such that
$V_1\sbs V_2\sbs V_3$ then $V_2\in\Rem{\cf}{U_i}$, i.e.
$\Rem{\cf}{U_i}$ is convex.\label{convex}
\end{enumerate}
\end{prop}

Every partition of a set induces an equivalence relation on this set.
The members of the partition comprise the equivalence classes. Since a
splitting induces a partition, we denote the equivalence relation induced by a
splitting $\cf$ by $\sim_\cf$.

\begin{dfn}
Given a set of open subsets $\cg$, we define the relation $\sim'_\cg$
on $\ct$ with $V_1\sim'_\cg V_2$ if and only if $V_1\sbs U\eqv V_2\sbs
U$ for all $U\in\cg$.
\end{dfn}

We have the following

\begin{prop}
The relation $\sim'_\cg$ is an equivalence.
\end{prop}

\begin{prop}
Given a finite splitting $\cf$, $\sim'_\cf = \sim_\cf$ i.e. the
remainders of $\cf$ are the equivalence classes of $\sim'_\cf$.
\label{prop:equiv}\end{prop}

We state some useful facts about splittings.

\begin{prop}
If $\cg$ is a finite set of opens, then ${\sf Cl(\cg)}$,
its closure under intersection, yields a finite splitting for $\dar\cg$.
\label{prop:closinter}
\end{prop}

The last proposition enables us to give yet another characterization
of remainders: every family of points in a complete lattice
closed under arbitrary joins comprises a
{\em closure system}, i.e. a set of fixed points of a closure operator
of the lattice (cf. \cite{COMP}.) Here, the lattice is the poset of
the
opens of the topological space. If we restrict ourselves to a finite
number of fixed points then we just ask for a finite set of opens
closed under intersection i.e. Proposition~\ref{prop:closinter}.
Thus a closure operator in the lattice of the open subsets of a
topological space induces an equivalence relation, two opens being
equivalent if they have the same closure, and the equivalence
classes of this relation are just the remainders of the open subsets
which are fixed points  of the closure operator.
The maximum open in $\Rem{\cf}{U}$, i.e. $U$, can be taken as the representative
of the equivalence class which is the union of all open
sets belonging to $\Rem{\cf}{U}$.

We now introduce the notion of stability corresponding to what we
mean by ``a formula retains its truth value on a set of opens''.

\begin{dfn}
If $\cg$ is a set of opens then $\cg$ is {\em stable for
$\phi$}, if for all $x$, either $x,V\sat\phi$ for all $V\in\cg$,
or $x,V\sat\neg\phi$ for all $V\in\cg$, such that $x\in V$.
\end{dfn}

\begin{prop}
If $\cg_1$,$\cg_2$ are sets of opens then
\label{prop:rem}
\renewcommand{\theenumi}{\alph{enumi}}
\begin{enumerate}
\item if $\cg_1\sbs\cg_2$ and $\cg_2$ is stable for
$\phi$ then $\cg_1$ is stable for $\phi$ ,\label{subrem}
\item
if $\cg_1$ is stable for $\phi$ and $\cg$ is stable for $\chi$ then
$\cg_1\cap\cg_2$ is stable for $\phi\land\chi$.\label{interem}
\end{enumerate}
\end{prop}

\begin{dfn}
A finite splitting $\cf=\{U_1,\ldots,U_n\}$ is called
a {\em stable splitting for} $\phi$, if $\Rem{\cf}{U_i}$ is stable
for $\phi$ for all $U_i\in\cf$.
\end{dfn}

\begin{prop} If $\cf=\{U_1,\ldots,U_n\}$ is a stable splitting for
$\phi$, so is
$${\cf}'={\sf Cl}(\{U_0,U_1,\ldots,U_n\}),$$
where $U_0\in\dar\cf$.\label{prop:unfsp}  \end{prop}

The above proposition tells us that if there is a
finite stable splitting for a topology then there is a closure
operator with finitely many fixed points whose associated equivalence classes
are stable sets of open subsets.

Suppose that  $\cm=\langle X,\ct,i\rangle$ is a topological model for
$\cl$. Let $\cf_\cm$ be a family of subsets of $X$ generated as follows:
$i(A)\in\cf_\cm$ for all $A\in\sa$, if $S\in\cf_\cm$ then
$X-S\in\cf_\cm$, if $S,T\in\cf_\cm$ then $S\cap T\in\cf_\cm$, and if
$S\in\cf_\cm$ then $S^\circ\in\cf_\cm$ i.e. $\cf_\cm$ is the least set
containing $\{i(A)|A\in\sa\}$ and closed under complements,
intersections and interiors. Let $\cf^\circ_\cm$ be the set
$\{S^\circ|S\in\cf_\cm\}$. We have
$\cf^\circ_\cm=\cf_\cm\cap\ct$. The following is the main theorem of this
section.

\begin{theorem}[Partition Theorem]
Let $\cm=\langle X,\ct,i\rangle$ be a topological model. Then there exists a
a set $\{\cf^\psi\}_{\psi\in\cl}$
of finite stable splittings such that
\begin{enumerate}
\item $\cf^{\psi}\sbs\cf^\circ_\cm$ and $X\in\cf^{\psi}$,  for all $\psi\in\cl$,
\item if $U\in\cf^\psi$ then $U^\psi=\{x\in U|x,U\sat\psi\}\in\cf_\cm$, and
\item if $\phi$ is a
       subformula of $\psi$ then $\cf^{\phi}\sbs\cf^{\psi}$ and $\cf^{\psi}$
       is a finite stable splitting for $\phi$,
\end{enumerate}
where $\cf_\cm$, $\cf_\cm^\circ$ as above.
\label{thm:main}
\end{theorem}

\begin{prf}
By induction on the structure of the formula $\psi$. In
each step we take care to refine the partition of the induction
hypothesis. Rather long proof.
\end{prf}

Theorem~\ref{thm:main} gives us a great deal of intuition for
topological models. It describes in detail the expressible
part of the topolocical lattice for the completeness result as it
appears in Chapter~\ref{ch:complete} and paves the road for
the reduction of the theory of topological models to that of spatial
lattices and the decidability result of this chapter.


\subsection{Basis Model}

     Let $\ct$ be a topology on a set $X$ and $\cb$ a basis for $\ct$.
We denote satisfaction in the models $\langle X,\ct,i\/\rangle$ and
$\langle X,\cb,i\/\rangle$ by $\tsat$ and $\bsat$, respectively.
In the following proposition we prove that each equivalence class
under $\sim_{\cf}$ contains an element of  a basis closed under finite
unions.

\begin{prop}
     Let $(X,\ct)$ be a topological space, and let $\cb$ be a basis
for $\ct$ closed under finite unions. Let $\cf$ be any finite subset
of $\ct$. Then for all $V \in \cf $ and all $x\in V$, there is some $U
\in \cb$ with $x \in U \subseteq V$ and $U\in\Rem{\cf}{V}$.
\label{prop:basis}
\end{prop}

\begin{corol}
     Let $(X,\ct)$ be a topological space, $\cb$ a basis for $\ct$
closed under finite unions, $x\in X$ and $U\in\cb$. Then
$$x,U\tsat\phi\quad\Eqv\quad x,U\bsat\phi.$$
\label{corol:subsatm}
\end{corol}

We shall prove that a model based on a topological space
$\ct$ is equivalent to the one induced by any basis of $\ct$ which is
lattice. Observe that this enables us to reduce the theory of
topological spaces to that of spatial lattices and, therefore, to
answer the conjecture of \cite{MP} : a completeness theorem for
subset spaces which are lattices will extend to the smaller class of
topological spaces.

\begin{theorem}
    Let $(X,\ct)$ be a topological space and $\cb$ a basis for $\ct$
closed under finite unions. Let $\cm_1=\langle X,\ct,i\/\rangle$ and
$\cm_2=\langle X,\cb,i\/\rangle$ be the corresponding models. Then, for
all $\phi$, $$\cm_1\sat\phi\quad\Eqv\quad\cm_2\sat\phi.$$
\label{thm:basis}
\end{theorem}


\subsection{Finite Satisfiability}

 \begin{prop} Let $\langle X,\ct\/\rangle$ be a subset space.
Let $\cf$ be a finite stable splitting for a formula $\phi$
 and all its subformulae, and assume that $X\in \cf$.
 Then for all $U\in\cf$,
 all $x\in U$, and all subformulae $\psi$ of $\phi$,
 $x, U \sat_\ct \psi$ iff $x, U \sat_\cf \psi$.
\label{prop:equivalent-models}
 \end{prop}

Constructing the quotient of $\ct$ under $\sim_{\cf}$ is not adequate
for generating a finite model because there may still be an infinite
number of points. It turns out that we only need a finite number of
them.

Let $\cm=\langle X,\ct,i\/\rangle$ be a
topological model, and define an equivalence relation $\sim$ on $X$ by
$x\sim y$ iff

\begin{description}
\item{(a)} for all $U \in \ct$, $x\in U$ iff $y \in U$, and
\item{(b)} for all atomic $A$, $x\in i(A)$ iff $y \in i(A)$.
\end{description}

Further, denote by $x^*$ the equivalence class of $x$,
and let $X^* = \{x^* : x \in X\}$. For every $U\in\ct$ let  $U^*=\{x^*
: x\in X\}$, then $\ct^* = \{ U^* : U \in \ct \}$ is a topology on
$X^*$. Define a map $i^*$ from the atomic formulae to the powerset
of $X^*$ by $ i^*(A) = \{ x^* : x\in i(A) \} $. The entire model $\cm$
lifts to the model $\cm^* = \langle X^*,\ct^*, i^* \/\rangle $ in a
well-defined way.

\begin{lemma} For all $x$, $U$, and $\phi$,
$$x, U \sat_\cm \phi \qquad\mbox{iff} \qquad x^*, U^* \sat_{\cm^*} \phi\ .$$
\label{lemma:quotient}
\end{lemma}

\begin{prf}
By induction on $\phi$.
\end{prf}

\begin{theorem}
If $\phi$ is satisfied in any topological  space then $\phi$
is satisfied in a finite topological space. \label{thm:finiteness}

\end{theorem}

Observe that  the finite
topological space
is a quotient of the initial one under two equivalences. The one equivalence is
$\sim_\cf^\phi$
on the open subsets of the topological space, where $\cf^\phi$ is the
finite splitting corresponding to $\phi$ and its cardinality is a
function of the complexity of $\phi$. The other
equivalence is $\sim_X$ on the points of the topological space and
its number of equivalence classes is a function of the atomic formulae
appearing in $\phi$.
The following simple example shows how a topology is formed with the quotient
under these two equivalences

\noindent {\em Example:} Let $X$ be the interval $[0,1)$ of real line with the
the set
$$\ct\quad=\quad\{\emptyset\}\cup\{\ [0,\frac{1}{2^n}) \ |\
n=0,1,2,\ldots\ \}$$
as topology. Suppose that we have only one atomic formula, call it $A$, such
that $i(A)=\{0\}$. then it is easy to see that the model $\langle
X,\ct,i\rangle$ is equivalent
to the finite topological model $\langle X^*,\ct^*,i^*\rangle$, where
$$
\begin{array}{c}
X^*\quad=\quad\{\  x_1,x_2\ \},\\
\ct^*\quad=\quad\{\ \emptyset, \{x_1,x_2\}\ \}, \hbox{ and}\\
i(A)\quad=\quad\{\  x_1 \ \}.
\end{array}
$$
\medskip

  So the overall size of the (finite) topological
space is bounded by a function of the complexity of $\phi$. Thus if we
want to test if a given formula is invalid we have a finite number of
finite topological spaces where we have to test its validity. Thus we
have the following

\begin{theorem}
The theory of topological spaces is decidable.
\end{theorem}

Observe that the last two results apply for lattices of subsets by
Theorem~{\ref{thm:basis}}.

\section{Completeness for ${\bf MP}^*$}
\label{ch:complete}

Open subsets of a topological space were used in \cite{MP} and in the previous
section to provide motivation, intuition and finally semantics for $\MP$. But in
this chapter we shall show that the canonical model of $\MP$ is actually
a set of subsets closed under arbitrary intersection and finite union, i.e. the
{\em closed} subsets of a topological space. However, these results are not
contrary
to our intuition for the following reasons: the spatial character of this
logic remains untouched. The fact that the canonical model is closed under
arbitrary intersections implies strong completeness with the much wider class  of
sets of subsets closed under finite intersection
and finite union. Now, the
results of the previous section allow us to deduce strong
completeness (in the sense that a consistent set of formulae is
simultaneously satisfiable
in  some model) also for the class of sets of subsets closed under infinite
union and finite intersection, i.e. the open subsets of a topological space.


\subsection{Subset frames}
\label{sec:closedframes}

As we noted in section~\ref{sec:lang},  we are not interpreting formulae
directly over a subset space but,
rather in the pointed product $X\ptimes\co$. The pointed product can be turned
in a set of possible worlds of a frame. We have only to indicate what the
accessibility relations are.

\begin{dfn}
Let $( X,\co )$ be a subset space. Its {\em subset frame} is the frame
$$\langle X\ptimes\co, \todam,\tol\rangle,$$
where
$$(x,U)\todam (y,V)\quad\hbox{if}\quad U=V$$
and
$$(x,U)\tol (y,V)\quad\hbox{if}\quad x=y \ \hbox{and}\ V\sbs U.$$
If $\co$ is a topology, intended as the closed subsets of a topological space,
we shall call its subset frame {\em closed topological frame}.
\end{dfn}

Our aim is to prove the most important properties of  such a frame.
We propose the following conditions on a possible worlds frame $\cf=\langle S,
R_1,R_2\rangle$ with two accessibility relations

\begin{enumerate}

\item \label{cond:po}
$R_1$ is reflexive and transitive.

\item \label{cond:equiv}
$R_2$ is an equivalence relation.

\item \label{cond:Barkan}
$R_1 R_2\sbs R_2 R_1$

\item \label{cond:end}
(ending points) $\cf$ has ending points with respect $R_1$, i.e
\begin{quote}
for all $s\in
S$ there exists $s_0\in S$ such that for all $s'\in S$ if $sR_1 s'$ then
$s'R_1 s_0$.
\end{quote}

\item \label{cond:ext}
(extensionality condition)
For all $s,s'\in S$, if there exists $s_0\in S$ such that $sR_1s_0$ and
$s'R_1s_0$ and
\begin{quote}
for all $t\in S$ such that  $tR_2 s$ there exist $t',t_0\in S$ such that
$t'R_2s'$, $tR_1t_0$
and $t'R_1t_0$, and  for all $t'\in S$ such that  $t'R_2 s'$ there exist
$t,t_0\in S$ such that $tR_2s$, $t'R_1t_0$ and $tR_1t_0$,
\end{quote}
then $s=s'$.

\item \label{cond:union}
(union condition)
For all $s_1,s_2\in S$,
\begin{quote}
if there exists $s\in S$ such that $sR_2R_1s_1$ and $sR_2R_1s_2$, then there
exists $s'\in S$ such that for all $t\in S$ with $tR_2s'$ then $tR_1R_2s_1$
or $tR_1R_2s_2$.
\end{quote}

\item \label{cond:inter}
(intersection condition) For all $\{s_i\}_{i\in I}\sbs S$,
\begin{quote}
if there exists $s\in S$ such that $s_iR_1s$ for all $i\in I$ then there exists
$s'\in S$ such that for all $\{t_i\}\sbs S$ with $t_iR_2s_i$ and
$t_iR_1t_0$ for all $i\in I$ and some $t_0\in S$ then $t_iR_1R_2 s'$.
\end{quote}

\item \label{cond:gen}
The frame $\cf$ is strongly generated in the sense that
\begin{quote}
there exists $s\in S$ such that for all $s'\in S$, $sR_2R_1s'$.
\end{quote}

\end{enumerate}

We have the following observations to make about the above conditions.
Conditions 1 to \ref{cond:union} and \ref{cond:gen} are first order, while
the  intersection condition is not. The extensionality condition
implies the following
\begin{quote}
for all $s,s'\in S$ such that $sR_1s_0$ and $s'R_2s_0$ then $s=s'$
\end{quote}
which implies that $R_1\cap R_2$ is the identity in $S$.
In view of the extensionality condition the relation $R_1$ is antisymmetric. So
we can replace condition~\ref{cond:po} with the condition that $R_1$ is a
partial order.

Now, we have the following proposition
\begin{prop}
If $(X,\ct)$ is a topological space then its closed topological frame
$\cf_{\ct}$ satisfies conditions \ref{cond:po} through~\ref{cond:gen}.
\end{prop}

The above proposition could lead to the consequence that topological models are
possible worlds models in disguise. But the following theorem shows that
this is not the case. There is a duality.

\begin{theorem}
\label{thm:isotopframe}
Let $\cf=\langle S,R_1,R_2\rangle$ be a frame satisfying conditions
\ref{cond:po} through~\ref{cond:gen}.
Then $\cf$ is isomorphic to a closed topological frame $\cf_\ct$.

\end{theorem}

Note that, in the above definitions, we could have used equally well the
equivalence class of $s\in S$ under the equivalence induced by the symmetric
closure of $R_1$ instead of the ending point of $s$ in $\cf$.
The above proofs show that the crucial conditions are conditions \ref{cond:po}
through~\ref{cond:ext} and if we are to strengthen or relax the union and
intersection conditions we get accordingly different conditions in the lattice
of the set of subsets of the space. The same holds for
condition~\ref{cond:gen}. We only used this condition to show that there exists
a top element, i.e. the whole space, and satisfy the hypothesis of the union
condition. If we do not assume this condition the union of two subsets will
belong to the set of subsets if they have an upper bound in it. We state this
case formally without a proof  because we are going to use it later.

\begin{prop}
\label{prop:conditionalunions}
\begin{enumerate}

\item Let $(X,\co)$ be a subset space closed under infinite intersections
and if $U,V\in\co$ have an upper bound in $\co$ then $U\cup V\in\co$. Then
its frame $\cf_{\co}$ satisfies conditions \ref{cond:po}
through~\ref{cond:inter}.

\item A frame $\cf$ satisfying conditions \ref{cond:po}
through~\ref{cond:inter} is isomorphic to a frame $\cf_{\co}$ where $(X,\co)$ as
in (1).

\end{enumerate}

\end{prop}


\subsection{On the proof theory of $\MP$}

We shall identify certain classes of formulae in $\cl$. This approach is
motivated by the results of Chapter~\ref{ch:sem}. In
fact, these formulae express definable parts of the lattice of subsets (see
section~\ref{sec:stability}.)

\begin{dfn}
Let $\cl'\sbs\cl$ be the set of formulae generated by the following
rules:

$$\sa\sbs\cl'\qquad\frac{\phi,\psi\in\cl'}{\phi\land\psi\in\cl'}\qquad
\frac{\phi\in\cl'}{\neg\phi,\dam\sk\phi\in\cl'}$$
Let $\cl''$ be the set $\{\sk\phi,\sfl\phi|\phi\in\cl'\}$.
\end{dfn}

Formulae in $\cl'$ have the following properties
\begin{dfn}
A formula $\phi$ of $\cl$ is called {\em persistent} whenever $\phi\to\Box\phi$
is a theorem (see also~\cite{MP}.)

A formula $\phi$ of $\cl$ is called {\em anti-persistent} whenever
$\neg\phi$ is persistent, i.e. $\neg\phi\to\Box\neg\phi$ (or, equivalently
$\dam\phi\to\phi$) is a theorem.

A formula $\phi$ of $\cl$ is called {\em bi-persistent} whenever
$\left(\phi\to\Box\phi\right)\land\left(\neg\phi\to\Box\neg\phi\right)$
(or, equivalently $\dam\phi\to\Box\phi$) is a
theorem.
\end{dfn}

Thus the truth of bi-persistent formulae depends only on the choice of the point
of the space while the satisfaction of persistent formulae can change at most
once in any model. We could go on and define a hierarchy of sets of formulae
where each member of hierarchy contains all formulae which their satisfaction
could change at most $n$ times in all models.

All the following derivations are in $\MP$
(Axioms 1 through \ref{ax:union} --- see table at page~\ref{table:mp}.)

\begin{prop}
All formulae belonging to $\cl'$ are bi-persistent.\label{prop:lpers}
\end{prop}

\begin{prf}
We prove it by induction, i.e. bi-persistence is retained through the
application of the formation rules of $\cl'$.
\end{prf}

A faster (semantical) proof would be ``the initial assignment on atomic formulae
extends to the wider class of $\cl'$''! This implies that formulae in $\cl'$
define subsets of the topological space.

Formulae in $\cl''$ have similar properties as the following lemma show.

\begin{lemma}
If $\phi$ is bi-persistent then
$\sk\phi$ is persistent and $\sfl\phi$ is anti-persistent.
\label{lem:boxkpers}
\end{lemma}

We prove some theorems of $\MP$ that we are going
to use later.

\begin{lemma} \label{lem:damand}
If $\phi$ is bi-persistent then
$\vdash_{\bf MP} \dam(\phi\land\psi)\equiv\dam\phi\land\dam\psi.$
\end{lemma}

The following is the key lemma to the DNF Theorem and generalizes
Axiom~\ref{ax:union}

\begin{lemma} \label{lem:main}
For all $n$,
$$\vdash_{\bf MP^*}
\dam\sk\phi\land\bland^n_{i=1}\sfl\left(\dam\sk\phi\land\psi_i\right)
\to\dam\left(\sk\phi\land\bland^n_{i=1}\sfl\psi_n\right),$$
where $\phi$, $\psi_i$ are bi-persistent.
\end{lemma}

All formulae of $\cl'$ can be expressed in terms of bi-persistent, persistent
and antipersistent formulae by means of the following normal form.

\begin{dfn}
\begin{enumerate}
\roman{enumi}
\item $\phi$ is in {\em prime normal} form (PNF) if it has the form
$$\psi\land\sk\psi'\land\bland^n_{i=1}\sfl\psi_i$$
where $\psi,\psi',\psi_i\in\cl'$ and $n$ is finite.
\item $\phi$ is in {\em disjunctive normal form} (DNF) if it has the
form $\blor^m_{i=1}\phi_i$, where each $\phi_i$ is in PNF and $m$ is
finite.
\end{enumerate}
\end{dfn}

We now give the formal analogue of the Partition Theorem.

\begin{theorem}[DNF]
For every $\phi\in\cl$, there is (effectively) a $\psi$ in DNF such
that
$$\vdash_{\bf MP^*}\phi\equiv\psi.$$
\label{thm:dnf}
\end{theorem}

The DNF theorem is the most important property of $\MP$. An immediate
corollary is that, as far as $\MP$ is concerned, we could have replaced the
$\Box$ modality with $\dam\sk$, since the formulae in normal form are defined
using these two modalities. Almost all subsequent
proof theoretic properties are immediate or implicit corollaries of the DNF
Theorem.

We close this section with the following proposition, which together with
Axiom~\ref{ax:direct} shows that $\Box\dam$ is equivalent to $\dam\Box$.

\begin{prop}
For all $\phi\in\cl$,
$\MPR\Box\dam\phi\to\dam\Box\phi$
\label{prop:boxdam}
\end{prop}


\subsection{Canonical Model}

The {\em canonical model} of $\bf MP^*$ is the structure
$$\clc=\left(S,\{\todam,\tol\},v\right),$$
where
$$\begin{array}{clc}
& S=\{s\sbs\cl|s \hbox{ is $\bf MP^*$-maximal consistent} \},& \\
& s\todam t \hbox{ iff }\{\phi\in\cl|\Box\phi\in s\}\sbs t, & \\
& s\tol t \hbox{ iff }\{\phi\in\cl|\sk\phi\in s\}\sbs t, & \\
& v(A)=\{s\in S | A\in S\}, &
\end{array}$$
along with the usual satisfaction relation (defined inductively):
$$\begin{array}{lll}
s\csat A & \hbox{iff} & s\in v(A)  \\
s\notcsat\bot & & \\
s\csat\neg\phi & \hbox{iff} & s\notcsat\phi  \\
s\csat\phi\land\psi & \hbox{iff} & s\csat\phi\hbox{ and }s\csat\psi \\
s\csat\Box\phi & \hbox{iff} & \hbox{for all }t\in S,\
                       s\todam t \hbox{ implies }t\csat\phi \\
s\csat\sk\phi & \hbox{iff} & \hbox{for all }t\in S,\
                       s\tol t \hbox{ implies }t\csat\phi.
\end{array}$$
We write $\clc\sat\phi$, if $s\csat\phi$ for all $s\in S$.

A canonical model exists for
all consistent bimodal systems with the normal axiom scheme for each
modality (as {\bf MP} and $\bf MP^*$.)
We have the following well known theorems (see~\cite{CH}, or~\cite{GL}.)

\begin{theorem}[Truth Theorem]
For all $s\in S$ and $\phi\in\cl$,
$$s\csat\phi\qquad\hbox{iff}\qquad\phi\in s.$$
\end{theorem}

\begin{theorem}[Completeness Theorem]
For all $\phi\in\cl$,
$$\clc\sat\phi\qquad\hbox{iff}\qquad\vdash_{\bf MP^*}\phi.$$
\end{theorem}

We shall now prove some properties of the members of $\clc$.
The DNF theorem implies that every maximal consistent theory $s$ of $\MP$ is
determined by the formulae in $\cl'$ and $\cl''$ it contains, i.e. by
$s\cap\cl'$ and $s\cap\cl''$.
Moreover, the set $\{\sk\phi,\sfl\phi|\sk\phi,\sfl\phi\in s\}$ is determined by
$s\cap\cl''$ alone (this is the $\sk$-case of the DNF theorem.)

The following definition is useful

\begin{dfn}
Let $P\sbs\cl'$. We say $P$ is an $\cl'$ theory if $P$ is consistent and for all
$\phi\in\cl'$ either $\phi\in P $ or $\neg\phi\in P$.

Let $S\sbs\cl''$. We say $S$ is an $\cl''$ theory if $S$ is consistent and for
all $\phi\in\cl''$ either $\phi\in S$ or $\neg\phi\in S$.
\end{dfn}

Hence, $s\cap\cl'$ is an $\cl'$ theory and $s\cap\cl''$ is an $\cl''$ theory.

What about going in the other direction? When does an $\cl'$ theory and
$\cl''$ theory determine an $\MP$ maximal consistent theory? When their union is
consistent because in this case there is a unique maximal extension.
To test consistency we have the following lemma.

\begin{lemma}
\label{lem:ltheories}
If $P$ and $S$ are an $\cl'$ and $\cl''$ theory respectively then $P\cup
S$ is consistent if and only if
$$ \hbox{if}\quad\phi\in P \quad\hbox{then}\quad\sfl\phi\in S.$$

\end{lemma}

It is expected that since $\cl'$ and $\cl''$ theories determine $\MP$ maximal
consistent sets they will determine their accessibility relations, as well.

\begin{prop} \label{prop:acchar}
For all $s,t\in S$,
$$\begin{array}{lll}
a.\ s\todam t & \hbox{if and only if} & \hbox{i. } \phi\in t
               \hbox{ if and only if } \phi \in s,
               \hbox{ where }\phi\in\cl', \\
     &             & \hbox{ii. if } \sfl\phi\in t \hbox { then }
                \sfl\phi \in s
                \hbox{, where }\phi,\psi\in\cl'. \\
b.\ s\tol t   & \hbox{if and only if} &
               \sk\phi\in t \hbox { if and only if } \sk\phi \in s,
               \hbox{ where } \phi\in\cl'.
\end{array}$$
\end{prop}

From the above proposition we have that
for all  $s,t\in S$, if
$s\todam t$ then $s\cap\cl'=t\cap\cl'$ and if $s\tol t$ then
$s\cap\cl''=t\cap\cl''$.

We write $\tol\todam$ for the composition of the relation $\tol$ and
$\todam$, i.e. if $s,t\in S$, we write $s\tol\todam t$ if there exists
$r\in S$ such that $s\tol r$ and $r\todam t$. Similarly for
$\todam\tol$.

For the composite relation $\tol\todam$ and $\todam\tol$ we have the following
corollary of proposition~\ref{prop:acchar}

\begin{corol}
\label{corol:cartodamtol}
For all $s,t\in S$,
$$\begin{array}{lll}
a.\ s\todam\tol t & \hbox{if and only if} & \hbox{i. if } \phi\in s
               \hbox{ then } \sfl\phi \in t,
               \hbox{ where }\phi\in\cl', \\
     &             & \hbox{ii. if } \sfl\phi\in t \hbox { then }
                \sfl\phi \in s
                \hbox{, where }\phi,\psi\in\cl'. \\
b.\ s\tol\todam t   & \hbox{if and only if} &
               \hbox{if } \sfl\phi\in t \hbox { then } \sfl\phi \in s,
               \hbox{ where } \phi\in\cl'.
\end{array}$$
\end{corol}

 We shall now prove that the canonical model $\clc$ of $\MP$ satisfies the
conditions of Section~\ref{sec:closedframes} on page~\pageref{sec:closedframes}.

We now have the following

\begin{theorem}
The canonical model $\clc$ of $\MP$ satisfies
conditions~1 to~7 of Section~\ref{sec:closedframes} on
page~\pageref{sec:closedframes}. \end{theorem}

\begin{corol}
\label{corol:canonicaltopology}
The canonical frame of $\MP$ is isomorphic to a subset  frame $\cf_{\co_c}$
where $(X_c,\co_c)$ is a subset space closed under infinite intersections
and if $U,V\in\co_c$ have an upper bound in $\co_c$ then $U\cup V\in\co_c$.
\end{corol}

By the construction of Theorem~\ref{thm:isotopframe}, $X_c$ consists of the
ending points of the members of the domain of the canonical model. We  define
the following initial assignment $i_c$
$$i(A)\quad=\quad\{\ s_0\ |\ A\in s_0\ \}.
$$
In this way the model $\cm=\langle X_c,\co_c,i_c\rangle$ is equivalent to the
canonical model as a corollary of frame
isomorphism.

\begin{corol}
For all $s\in S$ and $\phi\in\cl$ we have
$$
\phi\in s\quad \hbox{if and only if}\quad s_0,U_s\msat\phi.
$$
\end{corol}

\begin{prop}
The frame of a generated submodel $\clc^t$ is isomorphic to a closed topological
frame. \end{prop}

Now as above we have the following corollary

\begin{corol}
A submodel $\clc^t$ is equivalent to a closed topological model.
\end{corol}

It is a well known fact that a modal system is characterized by the class of
generated frames of the canonical frame.

\begin{prop}
\label{prop:strongclosed}
The system $\MP$ is (strongly) characterized by closed topological frames.

\end{prop}

Since the axioms and rules of $\MP$ are sound for the wider class of subset
spaces with finite union and intersection, we also have the following.

\begin{prop}
\label{prop:stronglattices}
The system $\MP$ is (strongly) characterized by subset frames closed under
finite unions and intersections.
\end{prop}

Now by Proposition~\ref{prop:strongclosed}
and~\ref{prop:stronglattices}, Corollary~\ref{corol:subsatm}
and Theorem~\ref{thm:basis} of Chapter~\ref{ch:sem}, where we proved the
equivalence of a topological model with the model induced by a basis closed
under finite unions, we have the following corollary

\begin{corol}
The system $\MP$ is (strongly) characterized by open topological frames as
well as subset frames closed under infinite unions and intersections.

\end{corol}

THe following {\em disjunction property} holds for $\MP$
$$\hbox{if}\qquad\MPR\sk\phi_1\lor\sk\phi_2\lor\ldots\lor\sk\phi_n\qquad
\hbox{then}\qquad\MPR\phi_i,\quad\hbox{for some }i,\quad 1\leq i\leq
n,$$
for $\phi_1,\phi_2,\ldots,\phi_n\in\cl'$.
Note that the disjunction rule does not hold for $\bf S5$.

\begin{prop}
$\MP$ provides the above rule of disjunction.

\end{prop}

We can similarly prove a stronger disjunction property, namely
$$
\begin{array}{ll}
\hbox{if} &
\MPR\sk\phi\to\sk\phi_1\lor\sk\phi_2\lor\ldots\lor\sk\phi_n\\
\hbox{then} & \MPR\sk\phi\to\phi_i,\quad\hbox{for some }i,\quad 1\leq
i\leq n,
\end{array}
$$
for $\phi,\phi_1,\phi_2,\ldots,\phi_n\in\cl'$.

Now we are able to prove the following
\begin{theorem}
The canonical model of $\MP$ is strongly generated.
\label{thm:gen}
\end{theorem}

By Theorem~\ref{thm:gen} we complete the set of conditions of
page~\pageref{cond:po} which turn the frame of the canonical model into a closed
subset frame. To summarize, we have the following corollary (note that the
canonical subset model is  $\langle X_c,\co_c,i_c\rangle$ of
Corollary~\ref{corol:canonicaltopology})

\begin{corol}
The canonical subset model of $\MP$ is a topological space. \end{corol}


\section{The Algebras of $\bf MP$ and $\MP$}
\label{ch:algebraic}

In this section we shall give a more general type of semantics for $\bf MP$ and
$\MP$. Every modal logic can be interpreted in an algebraic framework. An
algebraic model is nothing else but a valuation of the propositional variables
in a class of appropriately chosen  algebras.
We shall also make  connections with the previous chapters.

\subsection{ Fixed Monadic Algebras}

Interior operators were introduced by McKinsey and Tarski~\cite{MT}.

\begin{dfn}
An {\em interior} operator $I$ on a Boolean algebra $\cb=\langle B,{\bf 0},{\bf
1},\cap ,\cup \rangle $ is an operator satisfying  the conditions
$$
\begin{array}{c}
I(a\cap  b)=Ia\cap  Ib,\\
Ia\leq a,\\
IIa=Ia,\\
I{\bf 1}={\bf 1}.$$
\end{array}
$$

To each interior operator $I$ we associate its dual $C=-I-$, called {\em closure
operator}.\end{dfn}

Universal quantifiers were introduced by P. Halmos~\cite{HAL1}.

\begin{dfn}
A {\em universal quantifier} $\A $ on a Boolean algebra $\cb$ is an
operator satisfying  the conditions
$$
\begin{array}{c}
\A (a\cup  \A b)=\A a\cup  \A b,\\
\A a\leq a,\\
 \A {\bf 1}={\bf 1}.
\end{array}
$$

To each universal quantifier $\A $ we associate its dual $\E =-\A -$, called
{\em existential quantifier}. \end{dfn}

\begin{dfn}
Let $I$ be an interior operator on a Boolean algebra $\cb$. Let
$IB=\{a|a\leq Ia\}$ and $CB=\{a|Ca\leq a\}$, i.e. the fixed points of $I$ and
$C$ respectively. Let $B^{I}=IB\cap CB$ then $\cb^I=\langle B^I,{\bf 0},{\bf
1},-,\cap ,\cup \rangle $ is a Boolean subalgebra of $\cb$.
\end{dfn}

\begin{dfn}
A {\em fixed monadic algebra (FMA)} $\cb$ is a Boolean algebra with two
operators $I$ and $\A $ satisfying
$$\A I a\ge I\A a.$$

A valuation $v$ on $\cb$ is a function from the formulae of {\bf MP} to the
elements of $B$ such that
$$\begin{array}{rll}
 v(A)&\in &B^I,\hbox{ where $A$ is atomic},\\
 v(\neg\phi)&=&-v(\phi),\\
 v(\phi\land\psi)&=&v(\phi)\cap  v(\phi),\\
 v(\phi\lor\psi)&=&v(\phi)\cup  v(\phi),\\
 v(\Box\phi)&=&Iv(\phi),\\
 v(\sk\phi)&=&\A v(\phi).
\end{array}$$

An {\em algebraic model} of {\bf MP} is a FMA $\cb$ with a valuation $v$ on it.
We say $\phi$ is valid in this model iff $v(\phi)={\bf 1}$ and valid in an FMA
iff it is valid in all models based on this algebra. Finally, $\phi$ is
FMA-valid if it is valid in all FMA's. The notion of validity can extend to a
set of formulae.
\end{dfn}

Observe that the important part of the algebra is the smallest
subalgebra containing $B^I$ and closed under the operators $I$ and $\A $.

\begin{theorem}[Soundness for FMA-validity]
\label{thm:algsound}
If a formula $\phi$ is a theorem of {\bf MP} then $\phi$ is FMA-valid.
\end{theorem}

\begin{theorem}[Completeness for FMA-validity]
\label{thm:algcomplete}
If $\phi$ is FMA-valid then $\phi$ is a theorem of $\bf MP$.
\end{theorem}

\subsection{ Generated Monadic Algebras}

We shall now define the algebraic models of $\MP$

\begin{dfn}
A {\em generated monadic algebra (GMA)} $\cb$ is an FMA
satisfying in addition
$$
\begin{array}{c}
CIa=ICa\\
C(\A a\cap b)\cap \E C(\A a\cap c)\leq C(\A Ca\cap Cb\cap \E Cc).
\end{array}
$$
\end{dfn}

The concepts of algebraic model, validity, GMA-validity are defined as for
FMA's.
We used the direct algebraic translation of $\MP$ axioms but we could have
defined it with a different presentation. Observe that we only need $CIa\leq
ICa$ because the other direction is derivable (see
Proposition~\ref{prop:boxdam}.)

We now have the following

\begin{theorem}[Algebraic completeness of $\MP$]
A formula $\phi$ is a theorem of $\MP$ if and only if $\phi$ is GMA-valid.
\end{theorem}

It is known that a  modal algebra determines a (general) frame
(see~\cite{BS}.) So, in our case, the canonical algebraic model of $\MP$, i.e.
its Lindenbaum algebra, must determine a closed topological model (actually its
canonical frame.) We shall state only the interesting part of this
correspondence: the bijection on the domains. The accessibility relations are
defined in the usual way.

\begin{theorem}
There is a bijection between the set of the ultrafilters of the canonical
algebra of $\MP$ and the pointed product $X\ptimes\ct$, where $(X,\ct )$ is the
canonical topology of $\MP$.
\end{theorem}

The general theory of modal logic provides for yet another construction. A frame
determines a modal algebra. In case of the canonical frame, the modal algebra
determined must be isomorphic to the canonical modal algebra.
In our case, this algebra (which must be a GMA) has a nice representation. It is
the algebra of partitions of the topological lattice as it appeared in
Section~\ref{sec:stability}.

\section{Further Directions}
\label{ch:future}

There are several further directions
\begin{enumerate}
\item Due to the indeterminacy assumption (see Introduction) $\MP$ can be a
``core'' logical system for reasoning about computation with approximation or
uncertainty.

\item A discrete version of our epistemic framework can arise in
scientific experiments or tests. We acquire knowledge by ``a step-by-step''
process. Each step being an experiment or test. The outcome of such an
experiment or test is unknown to us beforehand, but after being known it
restricts our attention to a smaller set of possibilities. A sequence of
experiments, test or actions comprises a {\em strategy of knowledge
acquisition}. This model is in many respects similar to Hintikka's ``oracle''
(see~\cite{HI86}.) In Hintikka's model the ``inquirer'' asks a series of
questions to an external information source, called ``oracle''. The oracle
answers yes or no and the inquirer increases her knowledge by this piece of
additional evidence. This framework can be expressed by adding actions to the
language. Preliminary work of ours used {\em quantales} for modelling such
processes. A similar work without knowledge considerations appears in~\cite{AV}.

\item Since we can express concepts like affirmative or refutative assertions,
which are closed under infinite disjunctions and conjunctions respectively, it
is very natural to add infinitary connectives or fixed points operators (the
latter as a finite means to express the infinitary connectives.) This would
serve the purpose of specifying such properties of programs as ``emits an
infinite sequence of ones'' (see~\cite{AB91a} for a relevant discussion.)
An interesting direction of linking topological spaces with programs can be
found in~\cite{PAR83}.

\item Our work in the algebras of $\MP$ looks very promising. GMAs
(see~\ref{ch:algebraic}) have very interesting properties. A subalgebra of a GMA
corresponds to a complete space and this duality can be further investigated
with the algebraic machinery of modal logic (see~\cite{LEM66a},
\cite{LEM66b}, \cite{BL80}) or category theoretic methods.

\item Axiom~\ref{ax:boxk} forces monotonicity in our systems. If we drop this
axiom, an application of effort no longer implies a further increase in our
knowledge. Any change of our state of knowledge is possible. A non-monotonic
version of the systems presented in this thesis can be given along the lines
of~\cite{PAR87a}.

\item It would be interesting to consider a framework of multiple agents. Adding
a modality $\sk_i$ for each agent $i$ and assigning a different set of subsets
or topology to to each agent we can study their interaction or communication by
set-theoretic or topological means.

\item From our work became clear that both systems considered here are linked
with intuitionistic logic. We have embed intuitionistic logic to $\bf MP$ or
$\MP$  and it would be interesting to see how much of the
expressiveness of these logics can be carried in an intuitionistic framework.

\item Finally, in another direction Rohit Parikh considers an enrichment of
the language to express more (and purely) topological properties such as
separation properties and compactness.

\end{enumerate}

\newcommand{\etalchar}[1]{$^{#1}$}

\end{document}